\documentclass[aps, prl, reprint, superscriptaddress, amsmath, amssymb]{revtex4-1}
\usepackage{graphicx}%
\usepackage{dcolumn}%
\usepackage{bm}%
\usepackage{hyperref}%
\hypersetup{backref, colorlinks=true,linkcolor=blue, urlcolor=blue, citecolor=blue,linktocpage, linktoc=all}%
\usepackage{epstopdf}%
\usepackage{bbding}%
\draft

\usepackage{color}

\begin{document}
\title{Twisting enabled charge transfer excitons in epitaxially fused quantum dot molecules}%
\author{Yamei Zhou}
\affiliation{Key Laboratory for Special Functional Materials of Ministry of Education, Collaborative Innovation Center of Nano Functional Materials and Applications, and School of Materials Science and Engineering, Henan University, Kaifeng, Henan 475001, China}
\author{Christos S. Garoufalis}
\affiliation{Materials Science Department, University of Patras, 26504 Patras, Greece}
\author{Zaiping Zeng}
\email{zaiping.zeng@henu.edu.cn}
\affiliation{Key Laboratory for Special Functional Materials of Ministry of Education, Collaborative Innovation Center of Nano Functional Materials and Applications, and School of Materials Science and Engineering, Henan University, Kaifeng, Henan 475001, China}
\author{Sotirios Baskoutas}
\affiliation{Materials Science Department, University of Patras, 26504 Patras, Greece}
\author{Yu Jia}
\email{jiayu@henu.edu.cn}
\affiliation{Key Laboratory for Special Functional Materials of Ministry of Education, Collaborative Innovation Center of Nano Functional Materials and Applications, and School of Materials Science and Engineering, Henan University, Kaifeng, Henan 475001, China}
\affiliation{International Laboratory for Quantum Functional Materials of Henan, and School of Physics and Engineering, Zhengzhou University, Zhengzhou, Henan 450001, China}
\author{Zuliang Du}
\affiliation{Key Laboratory for Special Functional Materials of Ministry of Education, Collaborative Innovation Center of Nano Functional Materials and Applications, and School of Materials Science and Engineering, Henan University, Kaifeng, Henan 475001, China}
\date{\today}
\begin{abstract}
Charge-transfer excitons possessing long radiative lifetime and net permanent dipole moment are highly appealing for quantum dot (QD) based energy harvesting and photodetecting devices, in which the efficiency of charge separation after photo-excitation limits the device performance. Herein, using a hybrid time-dependent density functional theory, we have demonstrated that the prevailing rule of selecting materials with staggered type-II band alignment for realization of charge-transfer exciton breaks down in epitaxially fused QD molecules. The excitonic many-body effects are found to be significant and distinct depending on the exciton nature, causing unexpected reverse ordering of exciton states. Strikingly, twisting QD molecule appears as an effective means of modulating the orbital spatial localization towards charge separation that is mandatory for a charge-transfer exciton. Meanwhile, it manifests the intra-energy-level splitting that counterbalances the distinct many-body effects felt by excitons of different nature, thus ensuring the successful generation of energetically favourable charge-transfer exciton in both homodimer and heterodimer QD molecules. Our study extends the realm of twistroincs into zero-dimensional materials, and provides a genuine route of manipulating the exciton nature in QD molecules.
\end{abstract}
\maketitle

Colloidal semiconducting quantum dots (QDs) manifesting strong quantum confinement possess atomic-like characteristics with discrete electronic levels, which popularized the notation of QDs as artificial atoms. They are the key ingredients in many emerging technologies of light-electricity interconversion, including those of energy harvesting and photodetecting devices, owing to their high stability, and broad absorption spectral range\cite{Arquereaaz8541, Konstantatos180, Tang765, Won634, Shen192, Kim385}. In those devices, higher device performance requires efficient charge extraction at near-flat band, maximum-power conditions in which charge transport is diffusion-based rather than field-assisted. Because the diffusion length of minority carriers is shorter than the length required to maximize light absorption, this leads to an absorption-extraction compromise\cite{Arquereaaz8541}. A longterm challenge of possible extension of the diffusion length is the critical need of efficient charge separation after photo-excitation, therefore the formation of a charge-transfer (CT) exciton. Comparing with the highly localized Frenkel exciton with a short radiative lifetime, charge-transfer exciton with spatially separated electron-hole pair possesses a long radiative lifetime and a net permanent dipole moment\cite{Bardot035314, Rapaport033319}, which inhibits an efficient radiative charge recombination and thus is highly appealing for the light-electricity interconversion process. Despite the urgent demand of charge-transfer QD materials, it remains a longstanding puzzle how this important category of exciton can be practically achieved in QD systems.

\begin{figure*}[t!]
\centering
\includegraphics[scale=0.65]{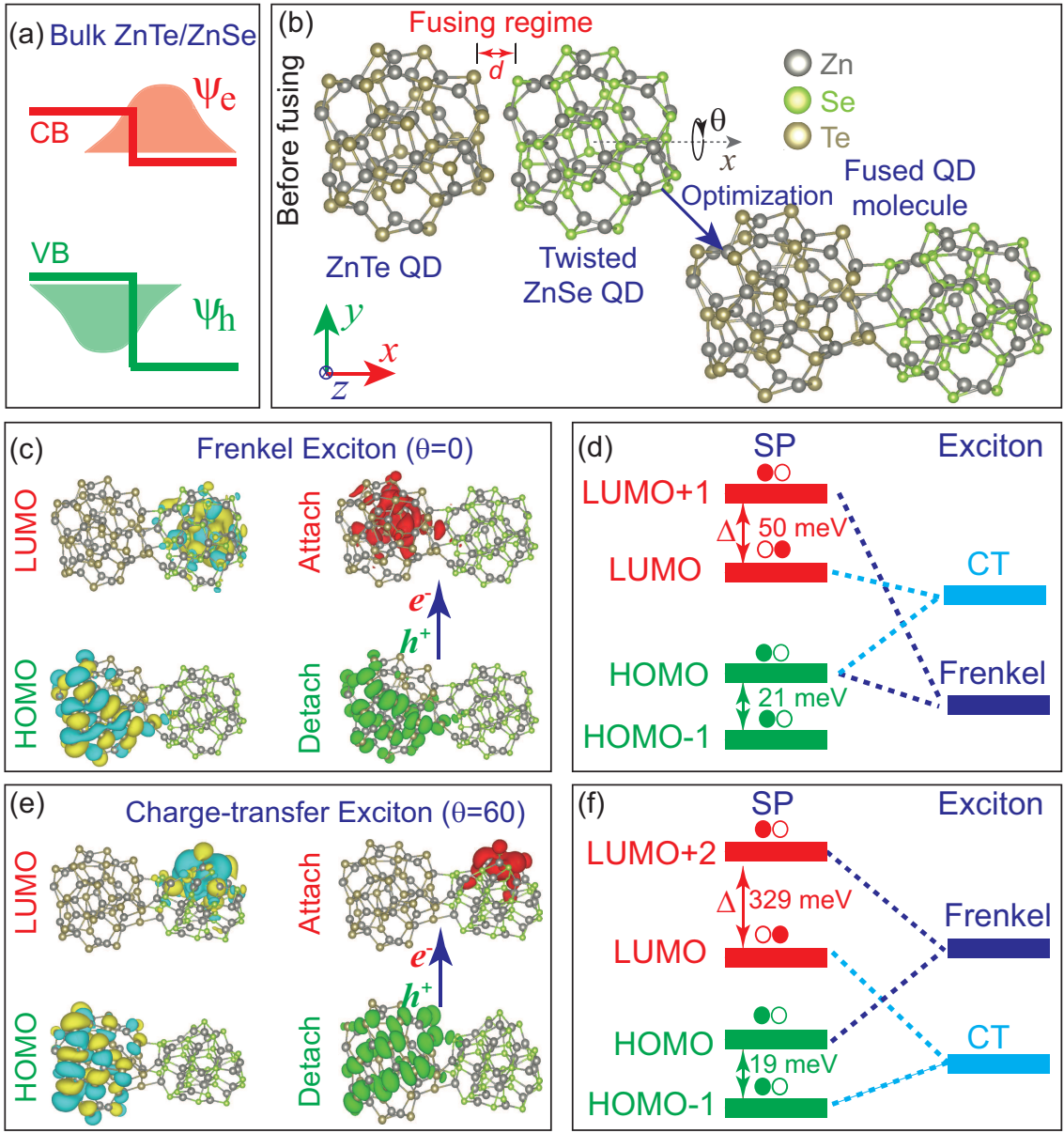}%
\caption{\label{fig1} (a) Schematic representation of conduction band (CB) and valence band (VB) alignment of bulk ZnTe/ZnSe heterojunction, and corresponding wave-function localization. (b) Epitaxially fused twisted ZnTe/ZnSe heterodimer QD molecule constructed by placing a ZnTe monomer QD and a twisted ZnSe monomer QD with a predefined twisted angle $\theta$ into fusing regime ($d<2.5 \ \rm \AA$), followed by a structural optimization to reach the lowest-energy geometric configuration. (c, e) Molecular orbital plots of HOMO and LUMO states, and attach and detach densities of the lowest-energy exciton in untwisted ((c), $\theta=0^{\circ}$) and twisted ((e), $\theta=60^{\circ}$) ZnTe/ZnSe heterodimer QD molecules. (d, f) The formation process of exciton manifold in untwisted ((c), $\theta=0^{\circ}$) and twisted ((e), $\theta=60^{\circ}$) ZnTe/ZnSe heterodimer QD molecules from the single-particle energy levels when considering many-body effects. The localization feature of the electron and hole states in monomer fragments of the molecule is schematically shown on top of each single-particle energy level, with open circle reflecting localization and filled circle meaning being empty, respectively. The intra-energy-level splitting $\Delta$ (e.g., conduction band herein) is documented alongside.}
\end{figure*}

\begin{figure*}[t!]
\centering
\includegraphics[scale=0.68]{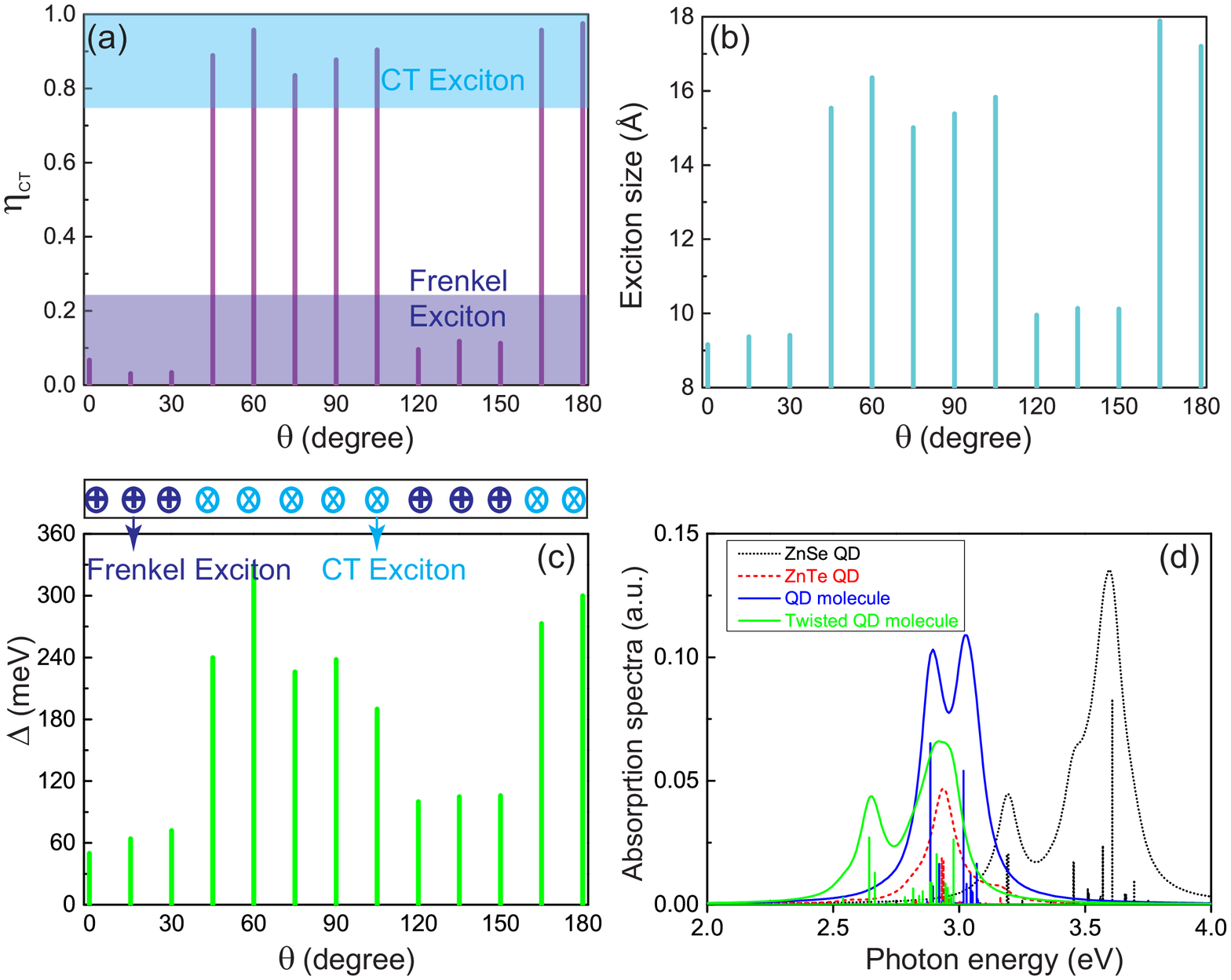}%
\caption{\label{fig2} (a) Charge-transfer index $\eta_{\rm CT}$, (b) exciton size, and (c) intra-conduction-band energy-level splitting $\Delta$ as a function of the twisted angle $\theta$ in twisted ZnTe/ZnSe QD molecule. In (c, d), the exciton nature, e.g., Frenkel exciton or charge-transfer exciton, is labeled on top of the plots. (d) Absorption spectrum of the untwisted ($\theta = 0^{\circ}$) and twisted ($\theta = 60^{\circ}$) ZnTe/ZnSe QD molecules compared with the corresponding spectrum of monomer ZnSe and ZnTe QDs.  Each spectra is obtained by calculating 30 optically bright singlet exciton states, and using a Lorentzian broadening function with broadening parameter $\Gamma$ = 0.05 eV. The vertical lines in each plot show the absorption peak corresponding to each exciton state.}
\end{figure*}

One of the geometric prerequisites for realizing charge-transfer exciton is the demand of heterostructured QD materials with a well-defined nanoscale interface. Colloidal QD dimer molecules formed by fusing QDs via oriented attachment, which are in analogy to diatomic molecules in atomic physics or inorganic chemistry, emerge as an unique platform that meets this requirement\cite{Cui5401, Koley1178, Salzmann787}. Besides inheriting all the superior optical properties of colloidal QDs, those epitaxially fused QD molecules exhibit manifested quantum coupling over 100 meV that is resolvable at room temperature\cite{Cui5401, Hughes4670}. This prominent feature makes colloidal QD molecule outperform the traditional coupled QDs grown via molecular beam epitaxy (MBE) with tiny electronic coupling ($\sim$ few meV) that confines their utility only to low-temperature operations in specialized cryogenic applications\cite{Kim223, Loss120}. The formation of colloidal QD molecule based on the QD building blocks will undoubtedly enrich the material database, and holds prodigious promise for a diverse emerging applications\cite{Koley1178}, given the richness and diversity of colloidal QDs. Another prerequisite for the generation of charge-transfer exciton is the spatially separated electron and hole wave functions at single-particle ground-state level. In this regard, selecting bulk semiconductor materials with staggered energy level alignment, which is typically referred as a type-II heterojunction, has commonly been used as a criterion for predicting charge separation. In this contribution, we have theoretically demonstrated that this zeroth-order-approximated selecting rule breaks down in the epitaxially fused QD molecules. However, twisting the QD molecule appears an enabling means of realizing charge-transfer exciton in both homodimer and heterodimer QD molecules composed by materials with well-established type II or quasi-type II band alignment.

We choose the Cd$_{33}$Se$_{33}$ QD with diameter $\sim 1.3$ nm as a reference monomer-QD\cite{Kilina7717}. This magic-sized QD represents one of the smallest CdSe QDs that have been experimentally identified with mass spectroscopy\cite{Kasuya99}. Magic-sized QDs have been widely chosen as model systems for the study of a variety of properties of QD, representing an excellent balance between the richness of physics and computational cost\cite{Kilina7717, Kilina6515, Trivedi2086, Lystrom892, Kasuya99, Pedersen733}. QDs of other material systems are realized by substituting the cation and anion atoms with the corresponding targeted counterparts, and then following up by a full geometry optimization. Using those monomer QDs as building blocks, the epitaxially fused QD molecule is constructed by placing two monomer-QDs into fusing regime, followed by a structural optimization to reach the lowest-energy geometric configuration (cf. Fig. \ref{fig1}). The geometry optimization is performed in the framework of density functional theory (DFT) with generalized gradient approximation (GGA) of Perdew-Burke-Ernzerhof (PBE) type\cite{Perdew3865, TURBOMOLE}. The hybrid nonlocal exchange-correlation functional of Becke and Lee, Yang and Parr (B3LYP \cite{Stephens11623}) and a basis set of double zeta quality (namely, the def2-SVP basis sets of the Karlsruhe group\cite{Weigend3297, Weigend1057}) have been further employed to calculate reliably the single-particle electronic structure. The excitonic optical properties are calculated on top of the B3LYP results, using the linear-response time-dependent DFT (TDDFT)\cite{Casida1995, Ullrich2012}. We have demonstrated that the employment of such a hybrid TDDFT scheme enables a quantitative prediction of excitonic properties of colloidal QDs of groups II-VI and III-VI comparing with experiments\cite{Ma245404, Han045404}. The obtained many-body excitonic wave functions are analyzed by employing a fragment-based excited-state analysis within a correlated electron-hole picture\cite{Plasser084108}. We naturally divide the epitaxially fused QD molecule into two individual sub-fragments, each of which contains a monomer-QD, and introduce charge transfer index $\eta_{\rm CT}$ to reflect the charge transfer character of an exciton state. The value of $\eta_{\rm CT}$ ranges from $\eta_{\rm CT} = 0$ characterizing an ideal Frenkel exciton to $\eta_{\rm CT} = 1$ referring to an ideal charge-transfer exciton. In reality, the exciton nature in the QD molecule is often of mixed nature, with the value of $\eta_{\rm CT}$ being between 0 and 1. We therefore tentatively categorize the exciton state with $\eta_{\rm CT} \leq 0.25$ as Frenkel exciton, and that with $\eta_{\rm CT} \geq 0.75$ as charge-transfer exciton. The approximate exciton size $\tilde{d}_{exc}$ of a given exciton state is calculated as the root-mean-square separation of the electron and hole in a point charge approximation\cite{Plasser084108}.

\begin{figure*}[t!]
\centering
\includegraphics[scale=0.54]{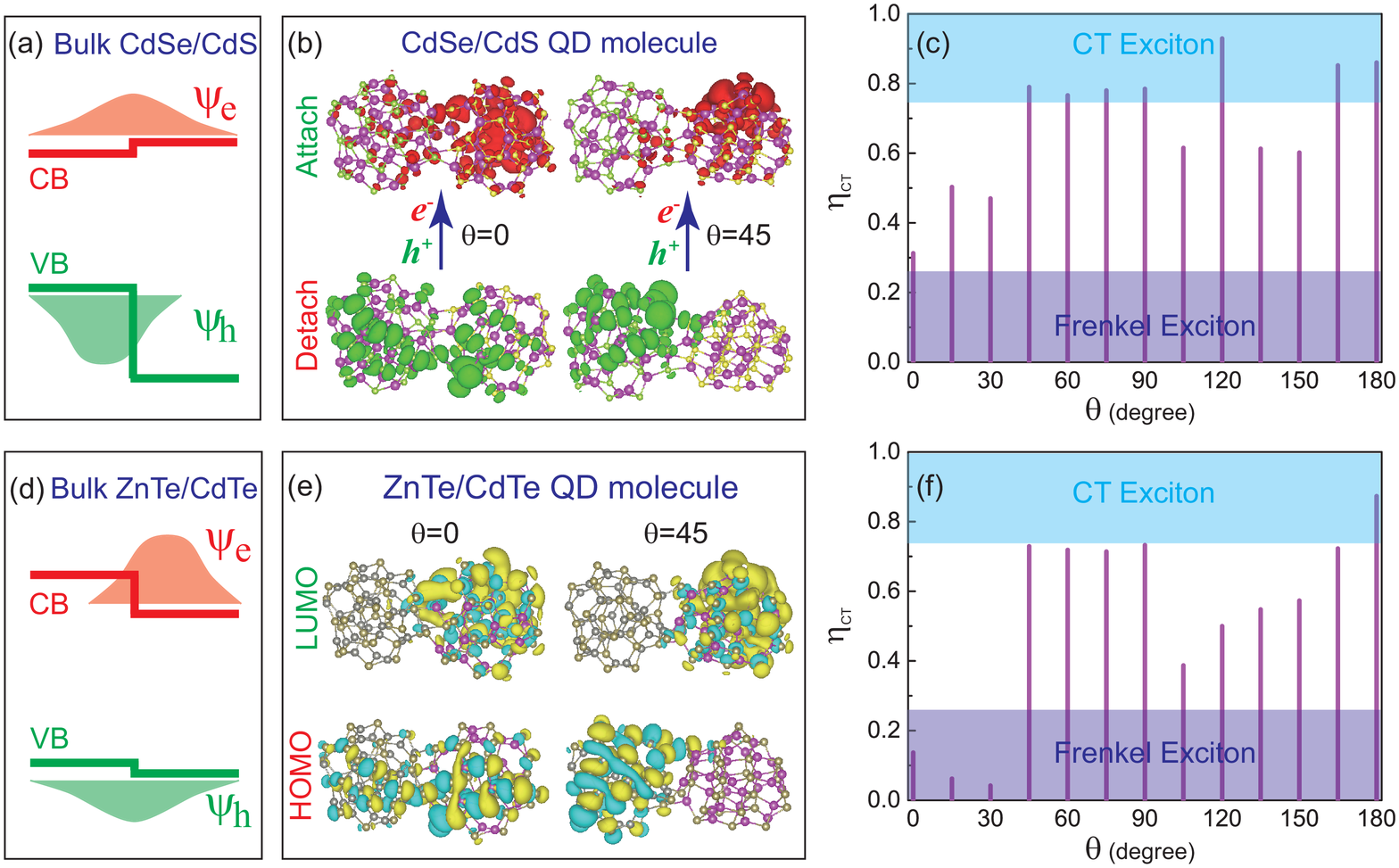}%
\caption{\label{fig3} (a, d) Schematic representation of conduction band (CB) and valence band (VB) alignment of bulk CdSe/CdS and ZnTe/CdTe heterojunctions, and corresponding wave-function localization. (b) The attach and detach densities of the lowest-energy exciton, and (e) the molecular orbital plots of HOMO and LUMO states in untwisted ($\theta=0^{\circ}$) and twisted ($\theta=45^{\circ}$) CdSe/CdS and ZnTe/CdTe QD molecules. (c, f) Charge-transfer index $\eta_{\rm CT}$ as a function of twisted angle $\theta$ in the twisted CdSe/CdS and ZnTe/CdTe QD molecules.}
\end{figure*}

\begin{figure*}[t!]
\centering
\includegraphics[scale=0.62]{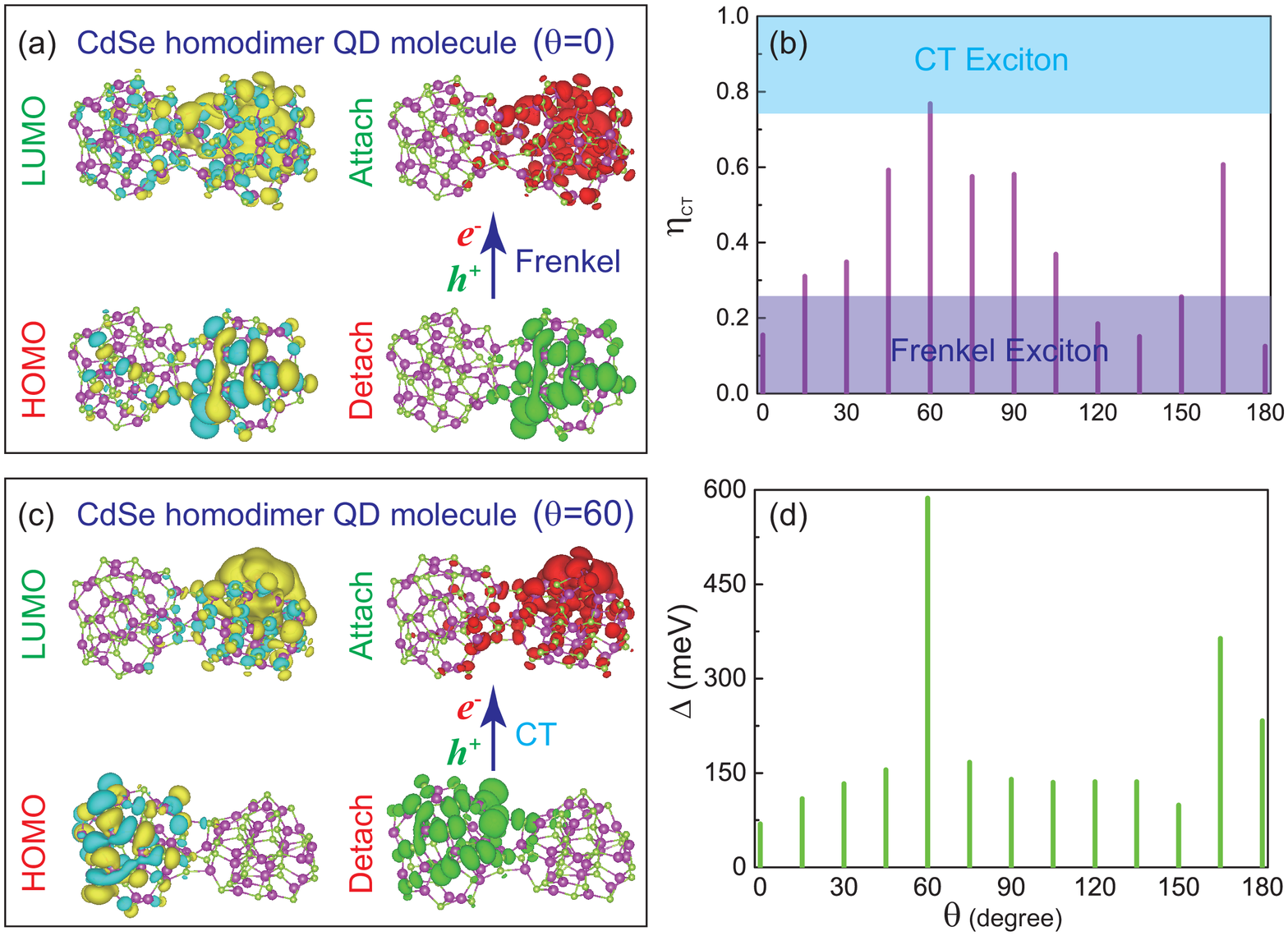}%
\caption{\label{fig4} (a, c) Molecular orbital plots of HOMO and LUMO states, and attach and detach densities of the lowest-energy exciton in untwisted ((a), $\theta=0^{\circ}$) and twisted ((c), $\theta=60^{\circ}$) CdSe homodimer QD molecules. (b) Charge-transfer index $\eta_{\rm CT}$ and (d) intra-valence-band energy-level splitting $\Delta$ as a function of the twisted angle $\theta$ in CdSe homodimer QD molecule.}
\end{figure*}

We first examine whether the prevailing rule of selecting heterostructured materials with staggered type II band alignment for realization of charge-transfer exciton holds in epitaxially fused QD molecules with well-defined nanoscale interface. We take the ZnTe/ZnSe heterodimer QD molecules as an example. Those material systems are known to exhibit a staggered interface when forming a bulk heterojunction (cf. Fig. \ref{fig1}(a)). We note that heterodimer molecules can already be synthesized in a controllable manner for both semiconducting QDs and metal nanoparticles using either partial anion exchange reaction\cite{Saruyama17598} or site-selected growth\cite{Qiu6703}. We find that the HOMO and LUMO states of the ZnTe/ZnSe QD molecule are spatially separately localized in ZnSe and ZnTe monomer QDs, respectively (cf. Fig. \ref{fig1}(c)), as expected. However, counterintuitively, when taking the many-body effects into account, the lowest optically bright exciton appears to be of Frenkel type rather than charge-transfer type. Similar picture has also been found in CdSe/CdTe QD molecule. The Frenkel exciton nature is evident from the corresponding attach and detach densities\cite{Martin14261}, which are both localized in the ZnTe monomer QD. We note that roughly speaking, the detach density is the denisty of the hole of a specific excitation and the attach density is the density of the excited electron. The Frenkel exciton nature is also characterized by the nearly vanishing charge-transfer index ($\eta_{\rm CT} = 0.067$; cf. Fig. \ref{fig2}(a)), and narrow exciton size ($\tilde{d}_{exc} =9.16 \ \rm \AA$; cf. Fig. \ref{fig2}(b)) being comparable to that of an isolated ZnTe QD ($\sim 9.42 \ \rm \AA$). The ground-state exciton manifold is found to primarily stem from the HOMO $\rightarrow \rm LUMO+1$ transition, rather than the HOMO $\rightarrow$ LUMO transition, as expected from the single-particle picture. This unconventional optics is attributed to the different Coulomb interaction between the electron and hole states involved in the transitions. Indeed, electron (LUMO+1) and hole (HOMO) states localized in the same fragment (e.g, Frenkel exciton) tend to exhibit a larger wave-function overlapping and feel more the many-body interaction (e.g., dominantly the Coulomb interaction) which drags the corresponding exciton manifold down to the ground state (cf. Fig. \ref{fig1}(d)). In stark contrast, the electron and hole states resided in spatially separated fragments exhibit a smaller wave-function overlapping and thus experience less the many-body effects (cf. Fig. \ref{fig1}(d)), which in turn causes an energetically-higher exciton manifold. This is corroborated by much larger exciton binding energy of the Frenkel exciton ($\sim$ 295 meV) than that of its charge-transfer counterpart ($\sim$ 183 meV).

In an attempt to possibly manipulate the nature of the band-edge exciton, we twist ZnTe/ZnSe QD molecule (cf. Fig. \ref{fig1}(b)). Consecutive rotation or twisting monomer QD is a natural process towards the epitaxially fused QD molecule via oriented attachment, which has been experimentally probed by in Situ Transmission Electron Microscopy (TEM)\cite{Marijn3959,Smeaton719}. Indeed, the twisted QD molecule exhibits comparable stability with its untwisted counterpart, as reflected by the marginal difference in the total energies ($<$ 0.2 eV). In stark contrast to the untwisted QD molecule ($\theta = 0^{\circ}$) having energetically favourable Frenkel exciton, ground-state charge-transfer exciton can be repeatedly realized in a twisted QD molecule while varying the twisted angle $\theta$ (cf. Fig. \ref{fig2}(a)). The nature of the charge-transfer exciton is characterized by the near-unity charge-transfer index $\eta_{\rm CT}$ (cf. Fig. \ref{fig2}(a)). This exciton manifold appears having a dominant contribution from the HOMO-1 $\rightarrow$ LUMO transition, in which both involved electronic states are spatially separately localized into two monomer QDs (cf. Fig. \ref{fig1}(f)). The charge-transfer exciton size is found to be significantly larger, being nearly twice of the Frenkel counterpart at a critical twisted angle $\theta = 165^{\circ}$ (cf. Fig. \ref{fig2}(b)). In order to explore the driving mechanism towards the twisting enabled charge-transfer exciton, we examine the intra-energy-level splitting $\Delta$ which is defined as the energetic difference between the lowest paired electronic states with respective orbital localization in spatially separated monomer QDs, which can be considered as an ``effective" band offset. $\Delta$ as a function of the twisted angle is shown in Fig. \ref{fig2}(c). Remarkably, it is found that the intra-conduction-band energy-level-splitting is drastically increased at the critical twisted angles whenever the ground-state charge-transfer exciton is reached (cf. Fig. \ref{fig2}(c)). For example, $\Delta$ reaches up to nearly 360 meV at critical twisted angle $\theta=60^{\circ}$ (e.g., charge-transfer exciton), which is nearly six times of the corresponding coupling strength at $\theta = 0^{\circ}$ ($\sim$ 60 meV; Frenkel exciton (cf. Fig. \ref{fig2}(c))). The twisting manifested intra-energy-level splitting therefore counterbalances the larger many-body interaction felt by the exciton manifolds involving the electron state (LUMO+2) at higher energy and the topmost hole state (e.g., HOMO-1 (cf. Fig. \ref{fig1}(f))) localized in the same monomer QD, resulting in a conventional optics with an energetically favourable charge-transfer exciton. Besides the exciton nature, twisting appears an enabling means of manipulating the excitonic characteristics of QD molecule, including red-shifting the absorption edge of the QD molecule and significantly reshaping the absorption spectra. Comparing with the untwisted QD molecule with $\theta=0^{\circ}$, the redshift reaches as much as 150 meV at twisted angle $\theta=60^{\circ}$  (cf. Fig. \ref{fig2}(d)).

To underpin the enabling power of twisting towards realization of charge-transfer exciton in epitaxially fused QD molecules, we further consider two heterostructured material systems with well-known quasi-type-II band alignment in bulk phase, (i) CdSe/CdS and (ii) ZnTe/CdTe (cf. Fig. \ref{fig3}(a,d)). In the former heterostructured materials, hole wave function is mainly localized in CdSe, while its electron counterpart is delocalized over both materials because of the tiny conduction band offset ($<$ 0.1 eV [Ref. \onlinecite{Hinuma155405}], cf. Fig. \ref{fig3}(a)), and vice versa for the latter one (cf. Fig. \ref{fig3}(d)). When going to the QD molecule with well-defined nanoscale interface, neither of those heterostructured material systems exhibits energetically favourable charge-transfer exciton (cf. Fig. \ref{fig3}(b, c, f)), as expected. For CdSe/CdS QD molecule, the lowest-energy exciton appears a hybrid Frenkel-charge-transfer exciton characterized by the attach and detach density plots in Fig. \ref{fig3}(b) ($\theta = 0^{\circ}$) and charge-transfer index $\eta_{\rm CT} = 0.32$ (cf. Fig. \ref{fig3}(c)). The hybrid exciton is originated from a nearly equal contribution from the HOMO-2 $\rightarrow$ LUMO and HOMO $\rightarrow$ LUMO transitions. In both transitions, the involved hole states occupy opposite QD fragments, causing a highly delocalized detach density (cf. Fig. \ref{fig3}(b); $\theta = 0^{\circ}$). For ZnTe/CdTe QD molecule, the electron and hole localization resembles very much the corresponding bulk heterostructure (cf. Fig. \ref{fig3}(d, e)). The energetically favourable exciton is found to be of Frenkel type (cf. Fig. \ref{fig3}(f)), and stems nearly purely from the HOMO $\rightarrow$ LUMO transition. Strikingly, charge-transfer exciton can be repeatedly realized in both QD molecules via twisting (cf. Fig. \ref{fig3}(b-c, f)). The realized exciton is exclusively contributed from the HOMO $\rightarrow$ LUMO transition with electron and hole states being localized in separate fragments, irrespectively of the critical twisted angles at which the ground-state charge-transfer exciton is achieved (cf. Figs. \ref{fig3}(c,f)). Two important driving forces enabled by twisting QD towards the charge-transfer exciton are identified. The first is that twisting manifests the intra-band energy-level-splitting, which dominates over the many-body interaction, avoiding the energy crossing between excitons of different nature. This twisting enabled effect is very similar to the situation in the afore-discussed QD molecule with staggered type II band alignment. The second is that twisting modulates the wave-function spatial localization. Indeed, the HOMO state of ZnTe/CdTe QD molecule switches from a highly delocalized nature over both monomer QDs ($\theta = 0^{\circ}$) to a highly localized feature ($\theta = 45^{\circ}$) (cf. Fig. \ref{fig3}(e)), which benefits to the realization of an energetically favourable charge-transfer exciton. We note that the modulation of wave-function spatial localization via twisting is only feasible when the band-offset between the neighbouring materials is considerably small as the quasi-type II band alignment considered herein, and is not observed in the QD molecules with type-II band alignment and large band-offsets (e.g., ZnTe/ZnSe QD molecule).

To consolidate the viable role of twisting on modulating the orbital spatial localization, we consider a CdSe homodimer QD molecule formed by orientated attachment of two identical CdSe monomer QDs. Homodimer QD molecule can be regarded as an extreme case of vanishing band-offset, and can be well-synthesized using wet chemistry methods as in Refs. \onlinecite{Hughes4670, Ondry12322}. We find that both HOMO and LUMO states of the homodimer molecule exhibit an asymmetric spatial distribution (cf. Fig. \ref{fig4}(a)) because of the lowered translational symmetry after fusing. This is contrary to what is expected from the continuum model like effective-mass approximation (e.g., symmetric charge distribution), therefore highlighting the importance of accurate atomistic simulations. The exciton state contributed from the HOMO $\rightarrow$ LUMO transition is energetically favourable, and appears to be of Frenkel type (cf. Fig. \ref{fig4}(a, b)). However, such a Frenkel exciton can be switched to its charge-transfer counterpart via twisting in the QD molecule at a critical twisted angle $\theta=60^{\circ}$ (cf. Fig. \ref{fig4}(b, c)). The realization of charge-transfer exciton is once again attributed to the twisting enabled modulation of orbital localization. Indeed, the orbital of HOMO state is mainly localized in the right monomer QD ($\theta = 0^{\circ}$), and switches its localization to nearly purely in the opposite monomer QD at the critical twisted angle ($\theta = 60^{\circ}$; cf. Fig. \ref{fig4}(c)). The resulted ground-state charge-transfer exciton appears a mixed contribution from the HOMO $\rightarrow$ LUMO and HOMO-1 $\rightarrow$ LUMO transitions, in which both involved hole states are exclusively localized in the left monomer QD. Remarkably, the intra-energy-level splitting $\Delta$ again peaks up at the critical twisted angle of $\theta = 60^{\circ}$, and appears nearly six times of that at $\theta=0^{\circ}$, therefore once again confirming the twisting manifested intra-energy-level splitting at the critical twisted angles (cf. Fig. \ref{fig4}(d)).

To summarize, we have shown that the prevailing guideline of selecting materials with staggered type-II band alignments for realization of energetically favourable charge-transfer exciton is not applicable in epitaxially fused QD molecules with strong quantum confinement effects and a nanoscale interface. The excitonic many-body effects are found to be significant and distinct depending on the exciton nature, causing unexpected exciton energy crossing. We have further demonstrated the enabling power of twisting on realization of ground-state charge-transfer exciton in homodimer and heterodimer QD molecules of material systems with staggered type II or quasi type-II band alignment. We have explored that twisting the QD molecules modulate the orbital spatial localization, and enables charge separation across the nanoscale interface that is mandatory for creating a charge-transfer exciton. Meanwhile, twisting enables a strong intra-energy-level splitting that counterbalances the distinct many-body effects felt by excitons of different nature, thus ensuring the generation of ground-state charge-transfer exciton. This study extends the realm of twistroincs into zero-dimensional material systems, and provides a clear guidance of manipulating the nature of band-edge exciton in QD molecules.\\
\\
The work has been partly supported by NSFC project with grant Nos. 11804077, 11774078, 12074099, and partly by the innovation research team of science and technology in Henan province (20IRTSTHN020). Z.Z. acknowledges the support of Distinguished Professor grant of Henan University with grant No. 2018001T.

\providecommand{\noopsort}[1]{}\providecommand{\singleletter}[1]{#1}%


\begin{thebibliography}{37}%
\makeatletter
\providecommand \@ifxundefined [1]{%
 \@ifx{#1\undefined}
}%
\providecommand \@ifnum [1]{%
 \ifnum #1\expandafter \@firstoftwo
 \else \expandafter \@secondoftwo
 \fi
}%
\providecommand \@ifx [1]{%
 \ifx #1\expandafter \@firstoftwo
 \else \expandafter \@secondoftwo
 \fi
}%
\providecommand \natexlab [1]{#1}%
\providecommand \enquote  [1]{``#1''}%
\providecommand \bibnamefont  [1]{#1}%
\providecommand \bibfnamefont [1]{#1}%
\providecommand \citenamefont [1]{#1}%
\providecommand \href@noop [0]{\@secondoftwo}%
\providecommand \href [0]{\begingroup \@sanitize@url \@href}%
\providecommand \@href[1]{\@@startlink{#1}\@@href}%
\providecommand \@@href[1]{\endgroup#1\@@endlink}%
\providecommand \@sanitize@url [0]{\catcode `\\12\catcode `\$12\catcode
  `\&12\catcode `\#12\catcode `\^12\catcode `\_12\catcode `\%12\relax}%
\providecommand \@@startlink[1]{}%
\providecommand \@@endlink[0]{}%
\providecommand \url  [0]{\begingroup\@sanitize@url \@url }%
\providecommand \@url [1]{\endgroup\@href {#1}{\urlprefix }}%
\providecommand \urlprefix  [0]{URL }%
\providecommand \Eprint [0]{\href }%
\providecommand \doibase [0]{http://dx.doi.org/}%
\providecommand \selectlanguage [0]{\@gobble}%
\providecommand \bibinfo  [0]{\@secondoftwo}%
\providecommand \bibfield  [0]{\@secondoftwo}%
\providecommand \translation [1]{[#1]}%
\providecommand \BibitemOpen [0]{}%
\providecommand \bibitemStop [0]{}%
\providecommand \bibitemNoStop [0]{.\EOS\space}%
\providecommand \EOS [0]{\spacefactor3000\relax}%
\providecommand \BibitemShut  [1]{\csname bibitem#1\endcsname}%
\let\auto@bib@innerbib\@empty
%</preamble>
\bibitem [{\citenamefont {Garc{\'\i}a~de Arquer}\ \emph
  {et~al.}(2021)\citenamefont {Garc{\'\i}a~de Arquer}, \citenamefont {Talapin},
  \citenamefont {Klimov}, \citenamefont {Arakawa}, \citenamefont {Bayer},\ and\
  \citenamefont {Sargent}}]{Arquereaaz8541}%
  \BibitemOpen
  \bibfield  {author} {\bibinfo {author} {\bibfnamefont {F.~P.}\ \bibnamefont
  {Garc{\'\i}a~de Arquer}}, \bibinfo {author} {\bibfnamefont {D.~V.}\
  \bibnamefont {Talapin}}, \bibinfo {author} {\bibfnamefont {V.~I.}\
  \bibnamefont {Klimov}}, \bibinfo {author} {\bibfnamefont {Y.}~\bibnamefont
  {Arakawa}}, \bibinfo {author} {\bibfnamefont {M.}~\bibnamefont {Bayer}}, \
  and\ \bibinfo {author} {\bibfnamefont {E.~H.}\ \bibnamefont {Sargent}},\
  }\href {https://science.sciencemag.org/content/373/6555/eaaz8541} {\bibfield
  {journal} {\bibinfo  {journal} {Science}\ }\textbf {\bibinfo {volume}
  {373}},\ \bibinfo {pages} {640} (\bibinfo {year} {2021})}\BibitemShut
  {NoStop}%
\bibitem [{\citenamefont {Konstantatos}\ \emph {et~al.}(2006)\citenamefont
  {Konstantatos}, \citenamefont {Howard}, \citenamefont {Fischer},
  \citenamefont {Hoogland}, \citenamefont {Clifford}, \citenamefont {Klem},
  \citenamefont {Levina},\ and\ \citenamefont {Sargent}}]{Konstantatos180}%
  \BibitemOpen
  \bibfield  {author} {\bibinfo {author} {\bibfnamefont {G.}~\bibnamefont
  {Konstantatos}}, \bibinfo {author} {\bibfnamefont {I.}~\bibnamefont
  {Howard}}, \bibinfo {author} {\bibfnamefont {A.}~\bibnamefont {Fischer}},
  \bibinfo {author} {\bibfnamefont {S.}~\bibnamefont {Hoogland}}, \bibinfo
  {author} {\bibfnamefont {J.}~\bibnamefont {Clifford}}, \bibinfo {author}
  {\bibfnamefont {E.}~\bibnamefont {Klem}}, \bibinfo {author} {\bibfnamefont
  {L.}~\bibnamefont {Levina}}, \ and\ \bibinfo {author} {\bibfnamefont {E.~H.}\
  \bibnamefont {Sargent}},\ }\href {\doibase 10.1038/nature04855} {\bibfield
  {journal} {\bibinfo  {journal} {Nature}\ }\textbf {\bibinfo {volume} {442}},\
  \bibinfo {pages} {180} (\bibinfo {year} {2006})}\BibitemShut {NoStop}%
\bibitem [{\citenamefont {Tang}\ \emph {et~al.}(2011)\citenamefont {Tang},
  \citenamefont {Kemp}, \citenamefont {Hoogland}, \citenamefont {Jeong},
  \citenamefont {Liu}, \citenamefont {Levina}, \citenamefont {Furukawa},
  \citenamefont {Wang}, \citenamefont {Debnath}, \citenamefont {Cha},
  \citenamefont {Chou}, \citenamefont {Fischer}, \citenamefont {Amassian},
  \citenamefont {Asbury},\ and\ \citenamefont {Sargent}}]{Tang765}%
  \BibitemOpen
  \bibfield  {author} {\bibinfo {author} {\bibfnamefont {J.}~\bibnamefont
  {Tang}}, \bibinfo {author} {\bibfnamefont {K.~W.}\ \bibnamefont {Kemp}},
  \bibinfo {author} {\bibfnamefont {S.}~\bibnamefont {Hoogland}}, \bibinfo
  {author} {\bibfnamefont {K.~S.}\ \bibnamefont {Jeong}}, \bibinfo {author}
  {\bibfnamefont {H.}~\bibnamefont {Liu}}, \bibinfo {author} {\bibfnamefont
  {L.}~\bibnamefont {Levina}}, \bibinfo {author} {\bibfnamefont
  {M.}~\bibnamefont {Furukawa}}, \bibinfo {author} {\bibfnamefont
  {X.}~\bibnamefont {Wang}}, \bibinfo {author} {\bibfnamefont {R.}~\bibnamefont
  {Debnath}}, \bibinfo {author} {\bibfnamefont {D.}~\bibnamefont {Cha}},
  \bibinfo {author} {\bibfnamefont {K.~W.}\ \bibnamefont {Chou}}, \bibinfo
  {author} {\bibfnamefont {A.}~\bibnamefont {Fischer}}, \bibinfo {author}
  {\bibfnamefont {A.}~\bibnamefont {Amassian}}, \bibinfo {author}
  {\bibfnamefont {J.~B.}\ \bibnamefont {Asbury}}, \ and\ \bibinfo {author}
  {\bibfnamefont {E.~H.}\ \bibnamefont {Sargent}},\ }\href {\doibase
  10.1038/nmat3118} {\bibfield  {journal} {\bibinfo  {journal} {Nat. Mat.}\
  }\textbf {\bibinfo {volume} {10}},\ \bibinfo {pages} {765} (\bibinfo {year}
  {2011})}\BibitemShut {NoStop}%
\bibitem [{\citenamefont {Won}\ \emph {et~al.}(2019)\citenamefont {Won},
  \citenamefont {Cho}, \citenamefont {Kim}, \citenamefont {Chung},
  \citenamefont {Kim}, \citenamefont {Chung}, \citenamefont {Jang},
  \citenamefont {Lee}, \citenamefont {Kim},\ and\ \citenamefont
  {Jang}}]{Won634}%
  \BibitemOpen
  \bibfield  {author} {\bibinfo {author} {\bibfnamefont {Y.-H.}\ \bibnamefont
  {Won}}, \bibinfo {author} {\bibfnamefont {O.}~\bibnamefont {Cho}}, \bibinfo
  {author} {\bibfnamefont {T.}~\bibnamefont {Kim}}, \bibinfo {author}
  {\bibfnamefont {D.-Y.}\ \bibnamefont {Chung}}, \bibinfo {author}
  {\bibfnamefont {T.}~\bibnamefont {Kim}}, \bibinfo {author} {\bibfnamefont
  {H.}~\bibnamefont {Chung}}, \bibinfo {author} {\bibfnamefont
  {H.}~\bibnamefont {Jang}}, \bibinfo {author} {\bibfnamefont {J.}~\bibnamefont
  {Lee}}, \bibinfo {author} {\bibfnamefont {D.}~\bibnamefont {Kim}}, \ and\
  \bibinfo {author} {\bibfnamefont {E.}~\bibnamefont {Jang}},\ }\href {\doibase
  10.1038/s41586-019-1771-5} {\bibfield  {journal} {\bibinfo  {journal}
  {Nature}\ }\textbf {\bibinfo {volume} {575}},\ \bibinfo {pages} {634}
  (\bibinfo {year} {2019})}\BibitemShut {NoStop}%
\bibitem [{\citenamefont {Shen}\ \emph {et~al.}(2019)\citenamefont {Shen},
  \citenamefont {Gao}, \citenamefont {Zhang}, \citenamefont {Lin},
  \citenamefont {Lin}, \citenamefont {Li}, \citenamefont {Chen}, \citenamefont
  {Zeng}, \citenamefont {Li}, \citenamefont {Jia}, \citenamefont {Wang},
  \citenamefont {Du}, \citenamefont {Li},\ and\ \citenamefont
  {Zhang}}]{Shen192}%
  \BibitemOpen
  \bibfield  {author} {\bibinfo {author} {\bibfnamefont {H.}~\bibnamefont
  {Shen}}, \bibinfo {author} {\bibfnamefont {Q.}~\bibnamefont {Gao}}, \bibinfo
  {author} {\bibfnamefont {Y.}~\bibnamefont {Zhang}}, \bibinfo {author}
  {\bibfnamefont {Y.}~\bibnamefont {Lin}}, \bibinfo {author} {\bibfnamefont
  {Q.}~\bibnamefont {Lin}}, \bibinfo {author} {\bibfnamefont {Z.}~\bibnamefont
  {Li}}, \bibinfo {author} {\bibfnamefont {L.}~\bibnamefont {Chen}}, \bibinfo
  {author} {\bibfnamefont {Z.}~\bibnamefont {Zeng}}, \bibinfo {author}
  {\bibfnamefont {X.}~\bibnamefont {Li}}, \bibinfo {author} {\bibfnamefont
  {Y.}~\bibnamefont {Jia}}, \bibinfo {author} {\bibfnamefont {S.}~\bibnamefont
  {Wang}}, \bibinfo {author} {\bibfnamefont {Z.}~\bibnamefont {Du}}, \bibinfo
  {author} {\bibfnamefont {L.~S.}\ \bibnamefont {Li}}, \ and\ \bibinfo {author}
  {\bibfnamefont {Z.}~\bibnamefont {Zhang}},\ }\href {\doibase
  10.1038/s41566-019-0364-z} {\bibfield  {journal} {\bibinfo  {journal} {Nature
  Photon.}\ }\textbf {\bibinfo {volume} {13}},\ \bibinfo {pages} {192}
  (\bibinfo {year} {2019})}\BibitemShut {NoStop}%
\bibitem [{\citenamefont {Kim}\ \emph {et~al.}(2020)\citenamefont {Kim},
  \citenamefont {Kim}, \citenamefont {Kim}, \citenamefont {Choi}, \citenamefont
  {Jang}, \citenamefont {Seo}, \citenamefont {Lee}, \citenamefont {Chung},\
  and\ \citenamefont {Jang}}]{Kim385}%
  \BibitemOpen
  \bibfield  {author} {\bibinfo {author} {\bibfnamefont {T.}~\bibnamefont
  {Kim}}, \bibinfo {author} {\bibfnamefont {K.-H.}\ \bibnamefont {Kim}},
  \bibinfo {author} {\bibfnamefont {S.}~\bibnamefont {Kim}}, \bibinfo {author}
  {\bibfnamefont {S.-M.}\ \bibnamefont {Choi}}, \bibinfo {author}
  {\bibfnamefont {H.}~\bibnamefont {Jang}}, \bibinfo {author} {\bibfnamefont
  {H.-K.}\ \bibnamefont {Seo}}, \bibinfo {author} {\bibfnamefont
  {H.}~\bibnamefont {Lee}}, \bibinfo {author} {\bibfnamefont {D.-Y.}\
  \bibnamefont {Chung}}, \ and\ \bibinfo {author} {\bibfnamefont
  {E.}~\bibnamefont {Jang}},\ }\href {\doibase 10.1038/s41586-020-2791-x}
  {\bibfield  {journal} {\bibinfo  {journal} {Nature}\ }\textbf {\bibinfo
  {volume} {586}},\ \bibinfo {pages} {385} (\bibinfo {year}
  {2020})}\BibitemShut {NoStop}%
\bibitem [{\citenamefont {Bardot}\ \emph {et~al.}(2005)\citenamefont {Bardot},
  \citenamefont {Schwab}, \citenamefont {Bayer}, \citenamefont {Fafard},
  \citenamefont {Wasilewski},\ and\ \citenamefont {Hawrylak}}]{Bardot035314}%
  \BibitemOpen
  \bibfield  {author} {\bibinfo {author} {\bibfnamefont {C.}~\bibnamefont
  {Bardot}}, \bibinfo {author} {\bibfnamefont {M.}~\bibnamefont {Schwab}},
  \bibinfo {author} {\bibfnamefont {M.}~\bibnamefont {Bayer}}, \bibinfo
  {author} {\bibfnamefont {S.}~\bibnamefont {Fafard}}, \bibinfo {author}
  {\bibfnamefont {Z.}~\bibnamefont {Wasilewski}}, \ and\ \bibinfo {author}
  {\bibfnamefont {P.}~\bibnamefont {Hawrylak}},\ }\href {\doibase
  10.1103/PhysRevB.72.035314} {\bibfield  {journal} {\bibinfo  {journal} {Phys.
  Rev. B}\ }\textbf {\bibinfo {volume} {72}},\ \bibinfo {pages} {035314}
  (\bibinfo {year} {2005})}\BibitemShut {NoStop}%
\bibitem [{\citenamefont {Rapaport}\ \emph {et~al.}(2006)\citenamefont
  {Rapaport}, \citenamefont {Chen},\ and\ \citenamefont
  {Simon}}]{Rapaport033319}%
  \BibitemOpen
  \bibfield  {author} {\bibinfo {author} {\bibfnamefont {R.}~\bibnamefont
  {Rapaport}}, \bibinfo {author} {\bibfnamefont {G.}~\bibnamefont {Chen}}, \
  and\ \bibinfo {author} {\bibfnamefont {S.~H.}\ \bibnamefont {Simon}},\ }\href
  {\doibase 10.1103/PhysRevB.73.033319} {\bibfield  {journal} {\bibinfo
  {journal} {Phys. Rev. B}\ }\textbf {\bibinfo {volume} {73}},\ \bibinfo
  {pages} {033319} (\bibinfo {year} {2006})}\BibitemShut {NoStop}%
\bibitem [{\citenamefont {Cui}\ \emph {et~al.}(2019)\citenamefont {Cui},
  \citenamefont {Panfil}, \citenamefont {Koley}, \citenamefont {Shamalia},
  \citenamefont {Waiskopf}, \citenamefont {Remennik}, \citenamefont {Popov},
  \citenamefont {Oded},\ and\ \citenamefont {Banin}}]{Cui5401}%
  \BibitemOpen
  \bibfield  {author} {\bibinfo {author} {\bibfnamefont {J.}~\bibnamefont
  {Cui}}, \bibinfo {author} {\bibfnamefont {Y.~E.}\ \bibnamefont {Panfil}},
  \bibinfo {author} {\bibfnamefont {S.}~\bibnamefont {Koley}}, \bibinfo
  {author} {\bibfnamefont {D.}~\bibnamefont {Shamalia}}, \bibinfo {author}
  {\bibfnamefont {N.}~\bibnamefont {Waiskopf}}, \bibinfo {author}
  {\bibfnamefont {S.}~\bibnamefont {Remennik}}, \bibinfo {author}
  {\bibfnamefont {I.}~\bibnamefont {Popov}}, \bibinfo {author} {\bibfnamefont
  {M.}~\bibnamefont {Oded}}, \ and\ \bibinfo {author} {\bibfnamefont
  {U.}~\bibnamefont {Banin}},\ }\href {\doibase 10.1038/s41467-019-13349-1}
  {\bibfield  {journal} {\bibinfo  {journal} {Nat. Commun.}\ }\textbf {\bibinfo
  {volume} {10}},\ \bibinfo {pages} {5401} (\bibinfo {year}
  {2019})}\BibitemShut {NoStop}%
\bibitem [{\citenamefont {Koley}\ \emph {et~al.}(2021)\citenamefont {Koley},
  \citenamefont {Cui}, \citenamefont {Panfil},\ and\ \citenamefont
  {Banin}}]{Koley1178}%
  \BibitemOpen
  \bibfield  {author} {\bibinfo {author} {\bibfnamefont {S.}~\bibnamefont
  {Koley}}, \bibinfo {author} {\bibfnamefont {J.}~\bibnamefont {Cui}}, \bibinfo
  {author} {\bibfnamefont {Y.~E.}\ \bibnamefont {Panfil}}, \ and\ \bibinfo
  {author} {\bibfnamefont {U.}~\bibnamefont {Banin}},\ }\href {\doibase
  10.1021/acs.accounts.0c00691} {\bibfield  {journal} {\bibinfo  {journal}
  {Acc. Chem. Res.}\ }\textbf {\bibinfo {volume} {54}},\ \bibinfo {pages}
  {1178} (\bibinfo {year} {2021})}\BibitemShut {NoStop}%
\bibitem [{\citenamefont {Salzmann}\ \emph {et~al.}(2021)\citenamefont
  {Salzmann}, \citenamefont {van~der Sluijs}, \citenamefont {Soligno},\ and\
  \citenamefont {Vanmaekelbergh}}]{Salzmann787}%
  \BibitemOpen
  \bibfield  {author} {\bibinfo {author} {\bibfnamefont {B.~B.~V.}\
  \bibnamefont {Salzmann}}, \bibinfo {author} {\bibfnamefont {M.~M.}\
  \bibnamefont {van~der Sluijs}}, \bibinfo {author} {\bibfnamefont
  {G.}~\bibnamefont {Soligno}}, \ and\ \bibinfo {author} {\bibfnamefont
  {D.}~\bibnamefont {Vanmaekelbergh}},\ }\href {\doibase
  10.1021/acs.accounts.0c00739} {\bibfield  {journal} {\bibinfo  {journal}
  {Acc. Chem. Res.}\ }\textbf {\bibinfo {volume} {54}},\ \bibinfo {pages} {787}
  (\bibinfo {year} {2021})}\BibitemShut {NoStop}%
\bibitem [{\citenamefont {Hughes}\ \emph {et~al.}(2014)\citenamefont {Hughes},
  \citenamefont {Blackburn}, \citenamefont {Kroupa}, \citenamefont {Shabaev},
  \citenamefont {Erwin}, \citenamefont {Efros}, \citenamefont {Nozik},
  \citenamefont {Luther},\ and\ \citenamefont {Beard}}]{Hughes4670}%
  \BibitemOpen
  \bibfield  {author} {\bibinfo {author} {\bibfnamefont {B.~K.}\ \bibnamefont
  {Hughes}}, \bibinfo {author} {\bibfnamefont {J.~L.}\ \bibnamefont
  {Blackburn}}, \bibinfo {author} {\bibfnamefont {D.}~\bibnamefont {Kroupa}},
  \bibinfo {author} {\bibfnamefont {A.}~\bibnamefont {Shabaev}}, \bibinfo
  {author} {\bibfnamefont {S.~C.}\ \bibnamefont {Erwin}}, \bibinfo {author}
  {\bibfnamefont {A.~L.}\ \bibnamefont {Efros}}, \bibinfo {author}
  {\bibfnamefont {A.~J.}\ \bibnamefont {Nozik}}, \bibinfo {author}
  {\bibfnamefont {J.~M.}\ \bibnamefont {Luther}}, \ and\ \bibinfo {author}
  {\bibfnamefont {M.~C.}\ \bibnamefont {Beard}},\ }\href {\doibase
  10.1021/ja413026h} {\bibfield  {journal} {\bibinfo  {journal} {J. Am. Chem.
  Soc.}\ }\textbf {\bibinfo {volume} {136}},\ \bibinfo {pages} {4670} (\bibinfo
  {year} {2014})}\BibitemShut {NoStop}%
\bibitem [{\citenamefont {Kim}\ \emph {et~al.}(2011)\citenamefont {Kim},
  \citenamefont {Carter}, \citenamefont {Greilich}, \citenamefont {Bracker},\
  and\ \citenamefont {Gammon}}]{Kim223}%
  \BibitemOpen
  \bibfield  {author} {\bibinfo {author} {\bibfnamefont {D.}~\bibnamefont
  {Kim}}, \bibinfo {author} {\bibfnamefont {S.~G.}\ \bibnamefont {Carter}},
  \bibinfo {author} {\bibfnamefont {A.}~\bibnamefont {Greilich}}, \bibinfo
  {author} {\bibfnamefont {A.~S.}\ \bibnamefont {Bracker}}, \ and\ \bibinfo
  {author} {\bibfnamefont {D.}~\bibnamefont {Gammon}},\ }\href {\doibase
  10.1038/nphys1863} {\bibfield  {journal} {\bibinfo  {journal} {Nat. Phys.}\
  }\textbf {\bibinfo {volume} {7}},\ \bibinfo {pages} {223} (\bibinfo {year}
  {2011})}\BibitemShut {NoStop}%
\bibitem [{\citenamefont {Loss}\ and\ \citenamefont
  {DiVincenzo}(1998)}]{Loss120}%
  \BibitemOpen
  \bibfield  {author} {\bibinfo {author} {\bibfnamefont {D.}~\bibnamefont
  {Loss}}\ and\ \bibinfo {author} {\bibfnamefont {D.~P.}\ \bibnamefont
  {DiVincenzo}},\ }\href {\doibase 10.1103/PhysRevA.57.120} {\bibfield
  {journal} {\bibinfo  {journal} {Phys. Rev. A}\ }\textbf {\bibinfo {volume}
  {57}},\ \bibinfo {pages} {120} (\bibinfo {year} {1998})}\BibitemShut
  {NoStop}%
\bibitem [{\citenamefont {Kilina}\ \emph {et~al.}(2009)\citenamefont {Kilina},
  \citenamefont {Ivanov},\ and\ \citenamefont {Tretiak}}]{Kilina7717}%
  \BibitemOpen
  \bibfield  {author} {\bibinfo {author} {\bibfnamefont {S.}~\bibnamefont
  {Kilina}}, \bibinfo {author} {\bibfnamefont {S.}~\bibnamefont {Ivanov}}, \
  and\ \bibinfo {author} {\bibfnamefont {S.}~\bibnamefont {Tretiak}},\ }\href
  {\doibase 10.1021/ja9005749} {\bibfield  {journal} {\bibinfo  {journal} {J.
  Am. Chem. Soc.}\ }\textbf {\bibinfo {volume} {131}},\ \bibinfo {pages} {7717}
  (\bibinfo {year} {2009})}\BibitemShut {NoStop}%
\bibitem [{\citenamefont {Kasuya}\ \emph {et~al.}(2004)\citenamefont {Kasuya},
  \citenamefont {Sivamohan}, \citenamefont {Barnakov}, \citenamefont {Dmitruk},
  \citenamefont {Nirasawa}, \citenamefont {Romanyuk}, \citenamefont {Kumar},
  \citenamefont {Mamykin}, \citenamefont {Tohji}, \citenamefont {Jeyadevan},
  \citenamefont {Shinoda}, \citenamefont {Kudo}, \citenamefont {Terasaki},
  \citenamefont {Liu}, \citenamefont {Belosludov}, \citenamefont
  {Sundararajan},\ and\ \citenamefont {Kawazoe}}]{Kasuya99}%
  \BibitemOpen
  \bibfield  {author} {\bibinfo {author} {\bibfnamefont {A.}~\bibnamefont
  {Kasuya}}, \bibinfo {author} {\bibfnamefont {R.}~\bibnamefont {Sivamohan}},
  \bibinfo {author} {\bibfnamefont {Y.~A.}\ \bibnamefont {Barnakov}}, \bibinfo
  {author} {\bibfnamefont {I.~M.}\ \bibnamefont {Dmitruk}}, \bibinfo {author}
  {\bibfnamefont {T.}~\bibnamefont {Nirasawa}}, \bibinfo {author}
  {\bibfnamefont {V.~R.}\ \bibnamefont {Romanyuk}}, \bibinfo {author}
  {\bibfnamefont {V.}~\bibnamefont {Kumar}}, \bibinfo {author} {\bibfnamefont
  {S.~V.}\ \bibnamefont {Mamykin}}, \bibinfo {author} {\bibfnamefont
  {K.}~\bibnamefont {Tohji}}, \bibinfo {author} {\bibfnamefont
  {B.}~\bibnamefont {Jeyadevan}}, \bibinfo {author} {\bibfnamefont
  {K.}~\bibnamefont {Shinoda}}, \bibinfo {author} {\bibfnamefont
  {T.}~\bibnamefont {Kudo}}, \bibinfo {author} {\bibfnamefont {O.}~\bibnamefont
  {Terasaki}}, \bibinfo {author} {\bibfnamefont {Z.}~\bibnamefont {Liu}},
  \bibinfo {author} {\bibfnamefont {R.~V.}\ \bibnamefont {Belosludov}},
  \bibinfo {author} {\bibfnamefont {V.}~\bibnamefont {Sundararajan}}, \ and\
  \bibinfo {author} {\bibfnamefont {Y.}~\bibnamefont {Kawazoe}},\ }\href
  {\doibase 10.1038/nmat1056} {\bibfield  {journal} {\bibinfo  {journal} {Nat.
  Mater.}\ }\textbf {\bibinfo {volume} {3}},\ \bibinfo {pages} {99} (\bibinfo
  {year} {2004})}\BibitemShut {NoStop}%
\bibitem [{\citenamefont {Kilina}\ \emph {et~al.}(2012)\citenamefont {Kilina},
  \citenamefont {Velizhanin}, \citenamefont {Ivanov}, \citenamefont {Prezhdo},\
  and\ \citenamefont {Tretiak}}]{Kilina6515}%
  \BibitemOpen
  \bibfield  {author} {\bibinfo {author} {\bibfnamefont {S.}~\bibnamefont
  {Kilina}}, \bibinfo {author} {\bibfnamefont {K.~A.}\ \bibnamefont
  {Velizhanin}}, \bibinfo {author} {\bibfnamefont {S.}~\bibnamefont {Ivanov}},
  \bibinfo {author} {\bibfnamefont {O.~V.}\ \bibnamefont {Prezhdo}}, \ and\
  \bibinfo {author} {\bibfnamefont {S.}~\bibnamefont {Tretiak}},\ }\href
  {\doibase 10.1021/nn302371q} {\bibfield  {journal} {\bibinfo  {journal} {ACS
  Nano}\ }\textbf {\bibinfo {volume} {6}},\ \bibinfo {pages} {6515} (\bibinfo
  {year} {2012})}\BibitemShut {NoStop}%
\bibitem [{\citenamefont {Trivedi}\ \emph {et~al.}(2015)\citenamefont
  {Trivedi}, \citenamefont {Wang},\ and\ \citenamefont
  {Prezhdo}}]{Trivedi2086}%
  \BibitemOpen
  \bibfield  {author} {\bibinfo {author} {\bibfnamefont {D.~J.}\ \bibnamefont
  {Trivedi}}, \bibinfo {author} {\bibfnamefont {L.}~\bibnamefont {Wang}}, \
  and\ \bibinfo {author} {\bibfnamefont {O.~V.}\ \bibnamefont {Prezhdo}},\
  }\href {\doibase 10.1021/nl504982k} {\bibfield  {journal} {\bibinfo
  {journal} {Nano Lett.}\ }\textbf {\bibinfo {volume} {15}},\ \bibinfo {pages}
  {2086} (\bibinfo {year} {2015})}\BibitemShut {NoStop}%
\bibitem [{\citenamefont {Lystrom}\ \emph {et~al.}(2021)\citenamefont
  {Lystrom}, \citenamefont {Roberts}, \citenamefont {Dandu},\ and\
  \citenamefont {Kilina}}]{Lystrom892}%
  \BibitemOpen
  \bibfield  {author} {\bibinfo {author} {\bibfnamefont {L.}~\bibnamefont
  {Lystrom}}, \bibinfo {author} {\bibfnamefont {A.}~\bibnamefont {Roberts}},
  \bibinfo {author} {\bibfnamefont {N.}~\bibnamefont {Dandu}}, \ and\ \bibinfo
  {author} {\bibfnamefont {S.}~\bibnamefont {Kilina}},\ }\href {\doibase
  10.1021/acs.chemmater.0c03610} {\bibfield  {journal} {\bibinfo  {journal}
  {Chem. Mater.}\ }\textbf {\bibinfo {volume} {33}},\ \bibinfo {pages} {892}
  (\bibinfo {year} {2021})}\BibitemShut {NoStop}%
\bibitem [{\citenamefont {Pedersen}\ \emph {et~al.}(1991)\citenamefont
  {Pedersen}, \citenamefont {Bjørnholm}, \citenamefont {Borggreen},
  \citenamefont {Hansen}, \citenamefont {Martin},\ and\ \citenamefont
  {Rasmussen}}]{Pedersen733}%
  \BibitemOpen
  \bibfield  {author} {\bibinfo {author} {\bibfnamefont {J.}~\bibnamefont
  {Pedersen}}, \bibinfo {author} {\bibfnamefont {S.}~\bibnamefont
  {Bjørnholm}}, \bibinfo {author} {\bibfnamefont {J.}~\bibnamefont
  {Borggreen}}, \bibinfo {author} {\bibfnamefont {K.}~\bibnamefont {Hansen}},
  \bibinfo {author} {\bibfnamefont {T.~P.}\ \bibnamefont {Martin}}, \ and\
  \bibinfo {author} {\bibfnamefont {H.~D.}\ \bibnamefont {Rasmussen}},\ }\href
  {\doibase 10.1038/353733a0} {\bibfield  {journal} {\bibinfo  {journal}
  {Nature}\ }\textbf {\bibinfo {volume} {353}},\ \bibinfo {pages} {733}
  (\bibinfo {year} {1991})}\BibitemShut {NoStop}%
\bibitem [{\citenamefont {Perdew}\ \emph {et~al.}(1996)\citenamefont {Perdew},
  \citenamefont {Burke},\ and\ \citenamefont {Ernzerhof}}]{Perdew3865}%
  \BibitemOpen
  \bibfield  {author} {\bibinfo {author} {\bibfnamefont {J.~P.}\ \bibnamefont
  {Perdew}}, \bibinfo {author} {\bibfnamefont {K.}~\bibnamefont {Burke}}, \
  and\ \bibinfo {author} {\bibfnamefont {M.}~\bibnamefont {Ernzerhof}},\ }\href
  {\doibase 10.1103/PhysRevLett.77.3865} {\bibfield  {journal} {\bibinfo
  {journal} {Phys. Rev. Lett.}\ }\textbf {\bibinfo {volume} {77}},\ \bibinfo
  {pages} {3865} (\bibinfo {year} {1996})}\BibitemShut {NoStop}%
\bibitem [{TUR()}]{TURBOMOLE}%
  \BibitemOpen
  \href@noop {} {\enquote {\bibinfo {title} {{TURBOMOLE V7.3 2018}, a
  development of {University of Karlsruhe} and {Forschungszentrum Karlsruhe
  GmbH}, 1989-2007, {TURBOMOLE GmbH}, since 2007; available from {\tt
  http://www.turbomole.com}.}}\ }\BibitemShut {NoStop}%
\bibitem [{\citenamefont {Stephens}\ \emph {et~al.}(1994)\citenamefont
  {Stephens}, \citenamefont {Devlin}, \citenamefont {Chabalowski},\ and\
  \citenamefont {Frisch}}]{Stephens11623}%
  \BibitemOpen
  \bibfield  {author} {\bibinfo {author} {\bibfnamefont {P.~J.}\ \bibnamefont
  {Stephens}}, \bibinfo {author} {\bibfnamefont {F.~J.}\ \bibnamefont
  {Devlin}}, \bibinfo {author} {\bibfnamefont {C.~F.}\ \bibnamefont
  {Chabalowski}}, \ and\ \bibinfo {author} {\bibfnamefont {M.~J.}\ \bibnamefont
  {Frisch}},\ }\href {\doibase 10.1021/j100096a001} {\bibfield  {journal}
  {\bibinfo  {journal} {J. Phys. Chem.}\ }\textbf {\bibinfo {volume} {98}},\
  \bibinfo {pages} {11623} (\bibinfo {year} {1994})}\BibitemShut {NoStop}%
\bibitem [{\citenamefont {Weigend}\ and\ \citenamefont
  {Ahlrichs}(2005)}]{Weigend3297}%
  \BibitemOpen
  \bibfield  {author} {\bibinfo {author} {\bibfnamefont {F.}~\bibnamefont
  {Weigend}}\ and\ \bibinfo {author} {\bibfnamefont {R.}~\bibnamefont
  {Ahlrichs}},\ }\href {\doibase 10.1039/B508541A} {\bibfield  {journal}
  {\bibinfo  {journal} {Phys. Chem. Chem. Phys.}\ }\textbf {\bibinfo {volume}
  {7}},\ \bibinfo {pages} {3297} (\bibinfo {year} {2005})}\BibitemShut
  {NoStop}%
\bibitem [{\citenamefont {Weigend}(2006)}]{Weigend1057}%
  \BibitemOpen
  \bibfield  {author} {\bibinfo {author} {\bibfnamefont {F.}~\bibnamefont
  {Weigend}},\ }\href {\doibase 10.1039/B515623H} {\bibfield  {journal}
  {\bibinfo  {journal} {Phys. Chem. Chem. Phys.}\ }\textbf {\bibinfo {volume}
  {8}},\ \bibinfo {pages} {1057} (\bibinfo {year} {2006})}\BibitemShut
  {NoStop}%
\bibitem [{\citenamefont {Casida}(1995)}]{Casida1995}%
  \BibitemOpen
  \bibfield  {author} {\bibinfo {author} {\bibfnamefont {M.~E.}\ \bibnamefont
  {Casida}},\ }\href@noop {} {\emph {\bibinfo {title} {Recent Advances in
  Density Functional Methods}}}\ (\bibinfo  {publisher} {World Scientific,
  Singapore},\ \bibinfo {year} {1995})\BibitemShut {NoStop}%
\bibitem [{\citenamefont {Ullrich}(2012)}]{Ullrich2012}%
  \BibitemOpen
  \bibfield  {author} {\bibinfo {author} {\bibfnamefont {C.}~\bibnamefont
  {Ullrich}},\ }\href@noop {} {\emph {\bibinfo {title} {Time-Dependent
  Density-Functional Theory: Concepts and Applications}}}\ (\bibinfo
  {publisher} {Oxford University Press},\ \bibinfo {year} {2012})\BibitemShut
  {NoStop}%
\bibitem [{\citenamefont {Ma}\ \emph {et~al.}(2019)\citenamefont {Ma},
  \citenamefont {Min}, \citenamefont {Zeng}, \citenamefont {Garoufalis},
  \citenamefont {Baskoutas}, \citenamefont {Jia},\ and\ \citenamefont
  {Du}}]{Ma245404}%
  \BibitemOpen
  \bibfield  {author} {\bibinfo {author} {\bibfnamefont {X.}~\bibnamefont
  {Ma}}, \bibinfo {author} {\bibfnamefont {J.}~\bibnamefont {Min}}, \bibinfo
  {author} {\bibfnamefont {Z.}~\bibnamefont {Zeng}}, \bibinfo {author}
  {\bibfnamefont {C.~S.}\ \bibnamefont {Garoufalis}}, \bibinfo {author}
  {\bibfnamefont {S.}~\bibnamefont {Baskoutas}}, \bibinfo {author}
  {\bibfnamefont {Y.}~\bibnamefont {Jia}}, \ and\ \bibinfo {author}
  {\bibfnamefont {Z.}~\bibnamefont {Du}},\ }\href {\doibase
  10.1103/PhysRevB.100.245404} {\bibfield  {journal} {\bibinfo  {journal}
  {Phys. Rev. B}\ }\textbf {\bibinfo {volume} {100}},\ \bibinfo {pages}
  {245404} (\bibinfo {year} {2019})}\BibitemShut {NoStop}%
\bibitem [{\citenamefont {Han}\ \emph {et~al.}(2021)\citenamefont {Han},
  \citenamefont {Min}, \citenamefont {Zeng}, \citenamefont {Garoufalis},
  \citenamefont {Baskoutas}, \citenamefont {Jia},\ and\ \citenamefont
  {Du}}]{Han045404}%
  \BibitemOpen
  \bibfield  {author} {\bibinfo {author} {\bibfnamefont {P.}~\bibnamefont
  {Han}}, \bibinfo {author} {\bibfnamefont {J.}~\bibnamefont {Min}}, \bibinfo
  {author} {\bibfnamefont {Z.}~\bibnamefont {Zeng}}, \bibinfo {author}
  {\bibfnamefont {C.~S.}\ \bibnamefont {Garoufalis}}, \bibinfo {author}
  {\bibfnamefont {S.}~\bibnamefont {Baskoutas}}, \bibinfo {author}
  {\bibfnamefont {Y.}~\bibnamefont {Jia}}, \ and\ \bibinfo {author}
  {\bibfnamefont {Z.}~\bibnamefont {Du}},\ }\href {\doibase
  10.1103/PhysRevB.104.045404} {\bibfield  {journal} {\bibinfo  {journal}
  {Phys. Rev. B}\ }\textbf {\bibinfo {volume} {104}},\ \bibinfo {pages}
  {045404} (\bibinfo {year} {2021})}\BibitemShut {NoStop}%
\bibitem [{\citenamefont {Plasser}(2020)}]{Plasser084108}%
  \BibitemOpen
  \bibfield  {author} {\bibinfo {author} {\bibfnamefont {F.}~\bibnamefont
  {Plasser}},\ }\href {\doibase 10.1063/1.5143076} {\bibfield  {journal}
  {\bibinfo  {journal} {J. Chem. Phys.}\ }\textbf {\bibinfo {volume} {152}},\
  \bibinfo {pages} {084108} (\bibinfo {year} {2020})}\BibitemShut {NoStop}%
\bibitem [{\citenamefont {Saruyama}\ \emph {et~al.}(2011)\citenamefont
  {Saruyama}, \citenamefont {So}, \citenamefont {Kimoto}, \citenamefont
  {Taguchi}, \citenamefont {Kanemitsu},\ and\ \citenamefont
  {Teranishi}}]{Saruyama17598}%
  \BibitemOpen
  \bibfield  {author} {\bibinfo {author} {\bibfnamefont {M.}~\bibnamefont
  {Saruyama}}, \bibinfo {author} {\bibfnamefont {Y.-G.}\ \bibnamefont {So}},
  \bibinfo {author} {\bibfnamefont {K.}~\bibnamefont {Kimoto}}, \bibinfo
  {author} {\bibfnamefont {S.}~\bibnamefont {Taguchi}}, \bibinfo {author}
  {\bibfnamefont {Y.}~\bibnamefont {Kanemitsu}}, \ and\ \bibinfo {author}
  {\bibfnamefont {T.}~\bibnamefont {Teranishi}},\ }\href {\doibase
  10.1021/ja2078224} {\bibfield  {journal} {\bibinfo  {journal} {J. Am. Chem.
  Soc.}\ }\textbf {\bibinfo {volume} {133}},\ \bibinfo {pages} {17598}
  (\bibinfo {year} {2011})}\BibitemShut {NoStop}%
\bibitem [{\citenamefont {Qiu}\ \emph {et~al.}(2019)\citenamefont {Qiu},
  \citenamefont {Xie}, \citenamefont {Lyu}, \citenamefont {Gilroy},
  \citenamefont {Liu},\ and\ \citenamefont {Xia}}]{Qiu6703}%
  \BibitemOpen
  \bibfield  {author} {\bibinfo {author} {\bibfnamefont {J.}~\bibnamefont
  {Qiu}}, \bibinfo {author} {\bibfnamefont {M.}~\bibnamefont {Xie}}, \bibinfo
  {author} {\bibfnamefont {Z.}~\bibnamefont {Lyu}}, \bibinfo {author}
  {\bibfnamefont {K.~D.}\ \bibnamefont {Gilroy}}, \bibinfo {author}
  {\bibfnamefont {H.}~\bibnamefont {Liu}}, \ and\ \bibinfo {author}
  {\bibfnamefont {Y.}~\bibnamefont {Xia}},\ }\href {\doibase
  10.1021/acs.nanolett.9b03167} {\bibfield  {journal} {\bibinfo  {journal}
  {Nano Lett.}\ }\textbf {\bibinfo {volume} {19}},\ \bibinfo {pages} {6703}
  (\bibinfo {year} {2019})}\BibitemShut {NoStop}%
\bibitem [{\citenamefont {Head-Gordon}\ \emph {et~al.}(1995)\citenamefont
  {Head-Gordon}, \citenamefont {Grana}, \citenamefont {Maurice},\ and\
  \citenamefont {White}}]{Martin14261}%
  \BibitemOpen
  \bibfield  {author} {\bibinfo {author} {\bibfnamefont {M.}~\bibnamefont
  {Head-Gordon}}, \bibinfo {author} {\bibfnamefont {A.~M.}\ \bibnamefont
  {Grana}}, \bibinfo {author} {\bibfnamefont {D.}~\bibnamefont {Maurice}}, \
  and\ \bibinfo {author} {\bibfnamefont {C.~A.}\ \bibnamefont {White}},\ }\href
  {\doibase 10.1021/j100039a012} {\bibfield  {journal} {\bibinfo  {journal} {J.
  Phys. Chem.}\ }\textbf {\bibinfo {volume} {99}},\ \bibinfo {pages} {14261}
  (\bibinfo {year} {1995})}\BibitemShut {NoStop}%
\bibitem [{\citenamefont {van Huis}\ \emph {et~al.}(2008)\citenamefont {van
  Huis}, \citenamefont {Kunneman}, \citenamefont {Overgaag}, \citenamefont
  {Xu}, \citenamefont {Pandraud}, \citenamefont {Zandbergen},\ and\
  \citenamefont {Vanmaekelbergh}}]{Marijn3959}%
  \BibitemOpen
  \bibfield  {author} {\bibinfo {author} {\bibfnamefont {M.~A.}\ \bibnamefont
  {van Huis}}, \bibinfo {author} {\bibfnamefont {L.~T.}\ \bibnamefont
  {Kunneman}}, \bibinfo {author} {\bibfnamefont {K.}~\bibnamefont {Overgaag}},
  \bibinfo {author} {\bibfnamefont {Q.}~\bibnamefont {Xu}}, \bibinfo {author}
  {\bibfnamefont {G.}~\bibnamefont {Pandraud}}, \bibinfo {author}
  {\bibfnamefont {H.~W.}\ \bibnamefont {Zandbergen}}, \ and\ \bibinfo {author}
  {\bibfnamefont {D.}~\bibnamefont {Vanmaekelbergh}},\ }\href {\doibase
  10.1021/nl8024467} {\bibfield  {journal} {\bibinfo  {journal} {Nano Lett.}\
  }\textbf {\bibinfo {volume} {8}},\ \bibinfo {pages} {3959} (\bibinfo {year}
  {2008})}\BibitemShut {NoStop}%
\bibitem [{\citenamefont {Smeaton}\ \emph {et~al.}(2021)\citenamefont
  {Smeaton}, \citenamefont {El~Baggari}, \citenamefont {Balazs}, \citenamefont
  {Hanrath},\ and\ \citenamefont {Kourkoutis}}]{Smeaton719}%
  \BibitemOpen
  \bibfield  {author} {\bibinfo {author} {\bibfnamefont {M.~A.}\ \bibnamefont
  {Smeaton}}, \bibinfo {author} {\bibfnamefont {I.}~\bibnamefont {El~Baggari}},
  \bibinfo {author} {\bibfnamefont {D.~M.}\ \bibnamefont {Balazs}}, \bibinfo
  {author} {\bibfnamefont {T.}~\bibnamefont {Hanrath}}, \ and\ \bibinfo
  {author} {\bibfnamefont {L.~F.}\ \bibnamefont {Kourkoutis}},\ }\href
  {\doibase 10.1021/acsnano.0c06990} {\bibfield  {journal} {\bibinfo  {journal}
  {ACS Nano}\ }\textbf {\bibinfo {volume} {15}},\ \bibinfo {pages} {719}
  (\bibinfo {year} {2021})}\BibitemShut {NoStop}%
\bibitem [{\citenamefont {Hinuma}\ \emph {et~al.}(2014)\citenamefont {Hinuma},
  \citenamefont {Gr\"uneis}, \citenamefont {Kresse},\ and\ \citenamefont
  {Oba}}]{Hinuma155405}%
  \BibitemOpen
  \bibfield  {author} {\bibinfo {author} {\bibfnamefont {Y.}~\bibnamefont
  {Hinuma}}, \bibinfo {author} {\bibfnamefont {A.}~\bibnamefont {Gr\"uneis}},
  \bibinfo {author} {\bibfnamefont {G.}~\bibnamefont {Kresse}}, \ and\ \bibinfo
  {author} {\bibfnamefont {F.}~\bibnamefont {Oba}},\ }\href {\doibase
  10.1103/PhysRevB.90.155405} {\bibfield  {journal} {\bibinfo  {journal} {Phys.
  Rev. B}\ }\textbf {\bibinfo {volume} {90}},\ \bibinfo {pages} {155405}
  (\bibinfo {year} {2014})}\BibitemShut {NoStop}%
\bibitem [{\citenamefont {Ondry}\ \emph {et~al.}(2019)\citenamefont {Ondry},
  \citenamefont {Philbin}, \citenamefont {Lostica}, \citenamefont {Rabani},\
  and\ \citenamefont {Alivisatos}}]{Ondry12322}%
  \BibitemOpen
  \bibfield  {author} {\bibinfo {author} {\bibfnamefont {J.~C.}\ \bibnamefont
  {Ondry}}, \bibinfo {author} {\bibfnamefont {J.~P.}\ \bibnamefont {Philbin}},
  \bibinfo {author} {\bibfnamefont {M.}~\bibnamefont {Lostica}}, \bibinfo
  {author} {\bibfnamefont {E.}~\bibnamefont {Rabani}}, \ and\ \bibinfo {author}
  {\bibfnamefont {A.~P.}\ \bibnamefont {Alivisatos}},\ }\href {\doibase
  10.1021/acsnano.9b03052} {\bibfield  {journal} {\bibinfo  {journal} {ACS
  Nano}\ }\textbf {\bibinfo {volume} {13}},\ \bibinfo {pages} {12322} (\bibinfo
  {year} {2019})}\BibitemShut {NoStop}%
\end{thebibliography}
\end{document}